# The Principle of the Malevolent Hiding Hand; or, the Planning Fallacy Writ Large

Bent Flyvbjerg and Cass R. Sunstein[*]

## Abstract

*We identify and document a new principle of economic behavior: the principle of the Malevolent Hiding Hand. In a famous discussion, Albert Hirschman celebrated the Hiding Hand, which he saw as a benevolent mechanism by which unrealistically optimistic planners embark on unexpectedly challenging plans, only to be rescued by human ingenuity, which they could not anticipate, but which ultimately led to success, principally in the form of unexpectedly high net benefits. Studying eleven projects, Hirschman suggested that the Hiding Hand is a general phenomenon. But the Benevolent Hiding Hand has an evil twin, the Malevolent Hiding Hand, which blinds excessively optimistic planners not only to unexpectedly high costs but also to unexpectedly low net benefits. Studying a much larger sample than Hirschman did, we find that the Malevolent Hiding Hand is common and that the phenomenon that Hirschman identified is rare. This sobering finding suggests that Hirschman's phenomenon is a special case; it attests to the pervasiveness of the planning fallacy, writ very large. One implication involves the continuing need for unbiased cost-benefit analyses and other economic decision support tools; another is that such tools might sometimes prove unreliable.*

## The Benevolent Hiding Hand

Albert O. Hirschman loved surprises, ironies, and paradoxes. He was delighted by human foibles and even more, he celebrated human creativity. He was fascinated by what he called "*petites idées*" and distrusted large claims and law-like generalizations, especially as a basis for policy. He enjoyed serendipity. He insisted that human history provides "stories, intricate and often nonrepeatable," which "look more like tricks history has up its sleeve than like social-scientific regularities, not to speak of laws." A refugee from Berlin during the rise of Nazism, he was keenly interested in "the many might-have-beens of history," including "felicitous and surprising escapes from disaster."

---

[*] Bent Flyvbjerg is the BT Professor and Chair of Major Programme Management at the University of Oxford; Cass Sunstein is Robert Walmsley University Professor at Harvard University. The authors are grateful to Dr. Dirk Bester for carrying out the statistical tests in the paper.

Despite his distrust of general social-scientific laws, Hirschman came up with quite a few large ideas of his own. One of these is the principle of the Hiding Hand, which is the cornerstone of his classic book, *Development Projects Observed*, first published in 1967 and recently reissued as a Brookings Classic (Hirschman 1967, 2015). The Hiding Hand turns out to be a bit of a trick up history's sleeve. It also provides a felicitous escape from disaster. It's a story, and an intricate one, but in Hirschman's view, it is repeatable. Hirschman believes that it tells us a great deal about development, if we are careful to specify the underlying mechanisms.

In Hirschman's account, social planners tend to be unrealistically optimistic, especially in underdeveloped nations. Ironically, that is fortunate, because if they were more realistic, they would not be bold enough to get started in the first place. Planners begin their projects by greatly overestimating some factor or condition that is indispensable to success, and underestimate difficulties and costs. According to Hirschman, planners thus tend to blunder in a predictable direction, because they neglect "a set of possible and unsuspected threats" to the profitability and even the ultimate existence of their projects. There is an evident connection here with the planning fallacy, much emphasized by behavioral scientists (Buehler 1994, Kahneman 2011), which suggests that people systematically underestimate the time that it will take to complete projects. To this point, Hirschman's argument can be seen as a version of the planning fallacy writ large (a claim to which we will return).

Fortunately, the planners' neglect of bad surprises is countered by a much happier surprise, which involves the sheer power of human creativity. Planners do not merely overestimate the likelihood of success and underestimate costs; they also underestimate potential responses to failure. Once things begin to go wrong, people discover unexpected ways to set them right, according to Hirschman. Hence the idea of a Hiding Hand, which "beneficially hides difficulties from us" and thus renders them invisible. The oddity is that while planners might never have authorized certain projects if they had had an accurate sense of the obstacles and costs that those projects would encounter, the result of the Hiding Hand is to produce an outcome that is as good as what the planner originally thought—or perhaps even better. This benevolent outcome is secured by what Hirschman called "providential ignorance" (Alacevich 2014: 157).

Hirschman offers two explanations for why planners tend to be blind to obstacles and challenges. He calls the first the "pseudo-imitation" technique, which means that planners pretend, or think, "that a project is nothing but a straightforward application of a well-known technique that has been successfully used elsewhere." The devastating problem, of course, is that situations and circumstances are different, so a project that is sold as if it were pure imitation usually has a large component of "indigenous initiative and execution."

The second explanation is the "pseudo-comprehensive-program" technique, by which planners dismiss previous efforts as piecemeal, and portray their own effort as a comprehensive program. With this technique, policymakers give, and are given, the illusion "that the 'experts' have already found all the answers," and all that is needed is faithful implementation. Those who enlist the pseudo-comprehensive-program technique end up underplaying the need for "imagination, insight, and the application of creative energies," thus covering up "the ignorance of the experts about the real cure of the malady they have been summoned to examine." We note that in the regulatory context, the pseudo-imitation technique and the pseudo-comprehensive-program technique both play a role, leading to projections of benefits and costs that are sometimes far too optimistic (Sunstein, 2013).

Sounding a bit like Friedrich Hayek, Hirschman argues that the two explanations turn out to work in concert, with the pseudo-imitation technique making "projects appear less difficulty-ridden than they actually are," and with the pseudo-comprehensive-program technique giving "project planners the illusion that they are in possession of far more insight into the projects' difficulties than is as yet available." Hence the decisionmaker has "crutches" that encourage him to proceed "at a stage when he has not yet acquired enough confidence in his problem-solving ability to make a more candid appraisal of a project's prospective difficulties and of the risks he is assuming." One of Hirschman's favorite examples was Pakistan's Karnaphuli paper mill, the biggest in Asia at the time of its construction (Hirschman 2015: 9-10). Examples from developed countries include Boston's Big Dig, the Channel tunnel connecting the UK and France, and Sydney's Opera House.

That's the bad news. The good news is that far from leading to disaster, the Hiding Hand provides both a spur and a remedy in the form of "a mechanism that makes the risk-averter take risks and in the process turns him into less of a risk-averter," according to Hirschman. Once risks are taken, human creativity emerges to solve unanticipated problems. And not only that. As decisionmakers become more experienced, they are able "to discard these crutches and to achieve a more mature appraisal of new projects." The crutches are a temporary learning device that helps bring risks down over time, "a transition mechanism through which decision makers learn to take risks, and *the shorter the transition and the faster the learning the better*" (Hirschman 2015: 28, 34; emphasis in original). "You have to do all these fool things before you do the sensible things," Hirschman observed in his field notes (Bianchi 2011: 18). The Hiding Hand is certainly not God, but it is pretty benevolent, and it works in mysterious ways. In fact, Hirschman invokes "Christianity's oft expressed preference for the repentant sinner," who has learned, "over the righteous man who never strays from the path of virtue." In the same vein, and even more on point, Hirschman invokes Nietzsche's maxim, "That which does not destroy me, makes me stronger."

Hirschman's focus is on development projects in poor nations, but he suggests, with evident delight and perhaps a touch of mischief, that "we may be dealing here with a fairly general phenomenon," which applies also in developed countries. That phenomenon has foundations in the claim that "people typically take on and plunge into new tasks because of the erroneously presumed absence of a challenge, because the task looks easier and more manageable than it will turn out to be." But upon finding that the underlying problems "are really more difficult than expected," and having been "stuck with them" because work is already well under way, people attack those problems ferociously—and often turn out to succeed. One might speculate, as has Offe (2013: 588), that Hirschman was referring, in part, to his own life experience, with his escape from Berlin, his extraordinary success in helping thousands of refugees to escape Nazi-occupied France, and his academic work.

Hirschman believed that notwithstanding its universality, the Hiding Hand is especially valuable in underdeveloped nations, "where confidence in creativity is lacking." In such nations, planners will be particularly reluctant to embark on projects that have a lot of unknowns, because their caution and lack of resources will make them unwilling to proceed. Their reluctance will be overcome only if they act on the basis of (misplaced, even blind) confidence that obstacles will not arise—and when they do, planners will be surprised by their power to surmount them, says Hirschman.

The challenge for close observers—and Hirschman was undoubtedly one—is to go beyond revealing anecdotes and interesting mechanisms to identify testable hypotheses. Can we test the idea of the Hiding Hand?

We might imagine productive exchanges between contemporary economists, insistent on large sample sizes and hypothesis testing, and Hirschman, with his distrust of universal laws. For conventional economists the questions are obvious: What, exactly, is Hirschman's hypothesis, and how do we know that it is true? Is he suggesting some generalization of the planning fallacy and offering also a hypothesis about its productive effects? We cannot entirely exclude the possibility that if pressed, Hirschman would respond that he is not, in fact, offering a testable hypothesis, but instead a description of a mechanism, and a widespread phenomenon, that greatly illuminates how projects actually operate, in developing and developed nations alike. The core of the mechanism is that people are "tricked"—in Hirschman's (2015:13) own word—by their ignorance of difficulties and costs into starting projects. But once the projects have been started, people find similarly underestimated sources of creativity to overcome and more than compensate for the initial difficulties and costs, making their projects succeed. In short, underestimated difficulties and costs at the outset are typically outweighed by even more underestimated problem-solving abilities and benefits during implementation, according to Hirschman's Benevolent Hiding Hand. And so Hirschman appeared to argue, in a way that suggests that he meant to describe a general pattern, and was not merely offering a mechanism and a series of tales.

## The Malevolent Hiding Hand

Hirschman is far too careful to insist that whenever planners overestimate the likelihood of success, the Hiding Hand will come to the rescue. He was well aware that projects exist, "from bankruptcies and white elephants to lost or ruinously won wars," for which underestimated difficulties and costs were not offset by even larger underestimates of creativity and benefits, but were instead exacerbated by visionlessness and benefit shortfalls (Hirschman 2015: 30). Thus the Benevolent Hiding Hand, which is Hirschman's topic, has an evil twin, the Malevolent Hiding Hand, which also hides obstacles and difficulties, but in situations in which creativity does not emerge, or emerges too late, or cannot possibly save the day. One of the fiendish acts of the Malevolent Hiding Hand is that it hides not only the initial obstacles and difficulties, but also the barriers to creativity itself. Indeed, Hirschman's own emphasis on the benevolence of the Hiding Hand might well serve (and perhaps has served) to empower the Malevolent Hiding Hand. For such situations, Streeten (1984: 116) talks about the "Principle of the Hiding Fist." Picciotto (1994b: 302) similarly observes that there might be "two hiding hands." One is applied by decisionmakers in the right situation, the other in the wrong one. In the latter case, "they [decisionmakers and their projects] would ... sink," concludes Picciotto.

The basic mechanisms driving the malevolent Hiding Hand are ignorance, psychology, and power (Flyvbjerg 2009). Ignorance points to the knowledge problem faced by planners of all kinds, even the most well-motivated (Hayek, 1945) and in particular to the difficulty of anticipated unintended consequences and systemic effects (Dorner, 1999). By itself, psychology might provide a sufficient explanation of the Malevolent Hiding Hand (as behavioral scientists have suggested, see Kahneman 2011). On that account, initial optimism is again an issue, but under the Malevolent Hiding Hand, such optimism applies to both the estimation of difficulties/costs and of creativity/benefits, whereas for the Benevolent Hiding Hand optimism applies to difficulties/costs but pessimism to creativity/benefits. For the Malevolent Hiding Hand, difficulties and costs therefore get optimistically underestimated, whereas creativity and benefits get just as optimistically overestimated. This double optimism at the outset comes back to haunt the project during delivery as a double whammy of cost overruns, delays, and other unanticipated hardships compounded by benefit shortfalls. Here is the planning fallacy writ very large, i.e., applying to not only schedule, and not only to initial obstacles (as Hirschman saw) but to costs and benefits in the widest sense. In the regulatory context, it is sometimes suggested, with confidence, that initial estimates of costs will turn out to be greatly overestimated, because regulated entities innovate and hence drive costs down. But sometimes such innovation does not occur, and initial estimates were actually far too optimistic (Sunstein, 2013).

Where optimism is innocent and unintentional, power play is deliberate and calculated. Planners who use power to activate the malevolent Hiding Hand deliberately underestimate difficulties/cost and overestimate creativity/benefits. They do this to make their projects look good on paper, which they see as increasing their chances of getting their projects approved and funded. Funding typically happens in tough competition with other projects in a tight budgetary process, leading to agency behavior and moral hazard. Advocates of regulation sometimes act similarly, underestimating costs and overestimating benefits in order to promote their goals. There might well be a thin line between deliberate underestimates/overestimates and *motivated reasoning* (Redlawsk, 2002), which can lead planners, and those who support their efforts, sincerely to believe in assessments that fit with their own hopes and commitments. We suspect that motivated reasoning often plays a large role and that there is a grey area between optimism and power where the two may blend and it is not always clear which is which, even to the actors involved.

It would be extravagant to insist that it is generally or universally good for planners to underestimate difficulties on the ground that people will discover inventive and unanticipated ways to solve those difficulties. Officials who are considering new regulations, knowing that compliance would be extremely expensive or impossible, ought not to proceed on the ground that technological innovation will inevitably make compliance inexpensive or feasible – even though impressive environmental innovation has sometimes occurred in the past. An uncharitable reading of Hirschman's work on the Hiding Hand would suggest that he has committed an identifiable error, which social scientists call "sampling on the dependent variable." Suppose, for example, that we wanted to understand what makes for a successful entrepreneur, and that we decided to find out by studying a set of successful entrepreneurs. Suppose we learned that the vast majority of them are exceedingly optimistic. From that finding, it would be a mistake to conclude that optimism is a necessary or sufficient condition for entrepreneurial success. There are a lot of failed entrepreneurs out there, and maybe most of them were exceedingly optimistic too. Maybe that trait, even if shared by the successful entrepreneurs, has no causal relationship to their success.

Hirschman identifies some striking instances of a Benevolent Hiding Hand, but his sample size is very small—only 11 projects—and his results are therefore open to random factors. There is little doubt that for countless unsuccessful development projects, the Hiding Hand did not work so well, or turned out to be malevolent, because the blindness at the initial stage is not countered by unanticipated creativity later on. True, and importantly, optimistic planners are sometimes rescued by such creativity, but much of the time, creativity is not triggered, and even if it is, it is not nearly enough to rescue their projects. In such instances, ignorance turned out not to be providential but inopportune instead. We could easily imagine an impressive if somewhat downbeat book, perhaps with the same title as Hirschman's classic, that catalogs a set of failures, bred by a failure to foresee obstacles and challenges that confound planners of many sorts. Indeed, it is not necessary to exercise our imaginations. James Scott's (1999) wild and brilliant book, *Seeing Like a State*, is merely the best example.

Hirschman did not, of course, produce such a book, and we do not believe that writing it would have much interested him. While he liked human foibles, he was delighted not by blunders and failures, but by history's generous tricks, by serendipity and silver linings, and perhaps above all by "felicitous and surprising escapes from disaster." He was no romantic, but he preferred happy endings, and he firmly believed that the Benevolent Hiding Hand "typically" applies, that is, in more cases than not, and that it was therefore the more overarching and more interesting principle for a general understanding of economic development and project behavior (Hirschman 2015: 1, 13). But with his small sample of 11 projects, Hirschman was in no position to establish whether this belief could be empirically substantiated, or which of the two Hiding Hands was the more prevalent. Hirschman glossed over the issue and basically seduced his readers into believing him, through storytelling and what Krugman (1994: 287) disapprovingly calls the "richness of plain English." (We suspect that like any good storyteller, Hirschman seduced not only his readers but also himself.)

Below, with a much larger sample than Hirschman's, we attempt to establish whether, and to what extent, Hirschman's principle of the Hiding Hand holds true and whether the Benevolent Hiding Hand or the Malevolent Hiding Hand is the more prevalent in policy and practice.

## Benevolent or Malevolent?

As mentioned, Hirschman developed the principle of the Hiding Hand based on data from a sample of 11 large projects. Here, we appraise the principle against a bigger and better dataset, from a sample of 2,062 large projects. We assess whether Hirschman's claims are supported by the findings from the greater sample. Such an assessment has not been done before.[1] The data are from the largest database of its kind, of estimated and actual costs and benefits in large projects. The data for the present study cover the eight project types listed in Table 1. They are infrastructure projects like most of the projects in Hirschman's smaller sample.[2] Geographically, the dataset includes projects in 104 countries on 6 continents, covering both developed and developing nations. Historically, the data cover almost a century, from 1927 to 2013. Older projects were included to enable analyses of historical trends. Data collection systematically followed international standards. Data collection and the database are described in detail in Flyvbjerg et al. (2002, 2005) and Flyvbjerg (2005).

Translating Hirschman's claims into a testable hypothesis is not as straightforward as it might seem; the translation requires a degree of extrapolation. We saw above that the core mechanism of Hirschman's principle of the Benevolent Hiding Hand is that people are tricked by their ignorance of difficulties and costs into starting projects, but once the projects have been started, people find similarly underestimated sources of creativity to overcome and more than compensate for the initial difficulties and costs, making their projects succeed. In this manner, underestimated difficulties and costs at the outset are outweighed by even more underestimated problem-solving abilities and benefits during implementation, resulting in net benefits and thus viable projects. If Hirschman's claims are right, we should therefore find in our dataset that higher-than-estimated project costs/difficulties are typically outweighed by even higher-than-estimated project benefits/problem-solving abilities. Consistent with his argument, we take him to have hypothesized that because of the Hiding Hand, the net benefits of plans end up as high as anticipated, or even higher, even if the costs turn out to be unexpectedly high. The high costs, and the difficulties that caused them, "set in motion a train of events that not only rescued the project but often made it particularly valuable," says Hirschman (2015: 12-13).

Table 1 shows our findings. The ideal test of Hirschman's hypothesis would catalogue costs and benefits (understood to capture all relevant factors) over the full life-cycle of a project. Unfortunately, such data are often unavailable. International convention is therefore to measure costs/difficulties by the proxy of construction costs and benefits/problem-solving abilities by the proxy of first-year benefits. This convention is followed here.[3] To be sure, first-year benefits may seem a narrow proxy to use for benefits. A possible objection to using them here is that the Benevolent Hiding Hand might emerge only after the first year, so that a focus on that will inevitably skew outcomes against Hirschman's claims. In fact, however, first-year benefits appear to be a highly reliable measure: For projects for which data are available on estimated and actual benefits covering more than one year after operations begin, it turns out that projects with lower-than-estimated benefits during the first year of operations also tend to have lower-than-estimated benefits in later years (Flyvbjerg 2013: 766-767). Using the first year as the basis for measuring benefits therefore does not appear to skew the analysis.

Cost overrun is measured as actual divided by estimated cost in real terms; benefit overrun is measured as actual divided by estimated usage, e.g., traffic for transportation infrastructure and power generation for energy infrastructure. For both costs and benefits, overrun is calculated with the baseline in the final business case, i.e., the date of the decision to build. Taking rail as an example, average cost overrun is listed in Table 1 as 1.40, which means that for rail projects actual costs turned out to be 40 percent higher than estimated costs on average and in real terms. Average benefit overrun for rail is listed as 0.66, which is evidence of a benefit shortfall of 34 percent, meaning that on average 34 percent of the estimated passengers never showed up on the actual trains.

If the basic idea of the Benevolent Hiding Hand were correct, average benefit overrun would be larger than average cost overrun. We see this is not the case for any of the eight project types in Table 1. Moreover, we see that for each and all project types on average there is no benefit overrun at all, but instead a benefit shortfall (benefit overrun < 1), which not only does not fit the Benevolent Hiding Hand, but runs diametrically counter to it.[4] From the p-values in Table 1 we see that the rejection of the Benevolent Hiding Hand applies at an overwhelmingly high level of statistical significance ($p<0.0001$, Mann-Whitney test), a level that is rarely found in studies of social phenomena.[5]

[Table 1 app. here]

To assess the robustness of results, we ran the same assessment for a subsample of 327 projects for which data were available for both cost overrun and benefit overrun for each project.[6] The results were similar. Average cost overrun for this subsample is 1.53, average benefit overrun 0.89 (compared with 1.39 and 0.9, respectively, for the larger sample). And again the Benevolent Hiding Hand is rejected at the same overwhelmingly high level of statistical significance ($p<0.0001$, paired Wilcoxon test).

Finally, we assessed results for the influence of project type and geography using Bayesian modeling.[7] We found only few significant differences across project type and geography—including between developing and developed nations—and none of them ran counter to the main conclusion above that higher-than-estimated costs are *not* outweighed by even higher-than-estimated benefits. This is unsurprising, given the overwhelmingly high level of statistical significance at which the main claim—benefit overruns outweigh cost overruns—was rejected.

It is particularly noteworthy that on average not only is benefit overrun not larger than cost overrun, as the Benevolent Hiding Hand says it would be, but on average there is *no benefit overrun at all*. Instead we find the opposite, namely a benefit shortfall. This shows that the idea of the Benevolent Hiding Hand is wrong both by degree and by direction as it gets both the size *and* the sign (plus instead of minus) wrong for benefit overrun. Instead of projects that generate benefits that compensate for cost overruns, as assumed by Hirschman (2015: 13) with his "two offsetting underestimates," in reality the average project is impaired by a double whammy of substantial cost overrun compounded by a substantial benefit shortfall. This is bad for viability, needless to say, and if projects are large enough and the economies where they are built are fragile, just one major project gone wrong can negatively affect the national economy for decades, as Brazil and Pakistan have learned with their large-dams projects (Ansar et al. 2014), and Greece with the 2004 Olympics (Flyvbjerg and Stewart 2012). This problem is not limited to public sector projects. Cost overruns, delays, and revenue shortfalls on the Airbus A380 jumbo jet put the company at risk and cost top management their jobs; K-Mart went out of business due to a billion-dollar IT project similarly gone wrong. Where is the Benevolent Hiding Hand when you need it, the owners of these projects might rightly have asked of Hirschman.

Table 2 answers the question of whether the Benevolent Hiding Hand is more common than its evil twin, the Malevolent Hiding Hand, as claimed by Hirschman. We here use the subsample of 327 projects for which data were available for both cost overrun and benefit overrun for each project. Again we see that Hirschman's claim is not supported. Not for a single one of the eight project types in Table 2 is the Benevolent Hiding Hand more common than the Malevolent Hiding Hand. On average, the Malevolent Hiding Hand dominates the Benevolent Hiding Hand by a factor 3.5 to 1. In other words, the prevalence of the Malevolent Hiding Hand is a full 255 percent higher than that of the Benevolent Hiding Hand. And again this result is supported at an overwhelmingly high level of statistical significance ($p<0.0001$, Mann-Whitney test, one-sided).

[Table 2 app. here]

Finally, we assessed Hirschman's (2015: 28) claim that risks will come down over time, because the Benevolent Hiding Hand is "essentially a transition mechanism through which decision makers learn to take risks." Schön (1994: 69) studied Hirschman's theory of learning and rightly observes that, "A theory about learning must deal with performance that improves over time. Performance that deteriorates, regresses, or merely swings from one mode of action to another does not qualify as learning." For the Benevolent Hiding Hand, improved performance would mean a reduction of project risks over time, resulting in cost overruns and benefit shortfalls coming down across projects over time, if not in the short term then in the medium and long run. If the data show such reduction, they support the Benevolent Hiding Hand on this point. If the data show no reduction, one would have to conclude that no learning takes place and that the Benevolent Hiding Hand does not apply. We used the data from Table 1 for which data also exist for opening year. The data cover 1,271 projects opened to service in the period from 1927 to 2011.

For costs, we found no significant relationship between cost overrun and time, that is, cost overrun neither decreased nor increased over time (BF= 1.70, Bayesian test).[8] For benefits, we found a statistically highly significant historical trend of declining benefit overruns (increasing benefit shortfalls) of 0.5 percent per year, which is the opposite of what Hirschman claimed for the Benevolent Hiding Hand (BF=799). This is when using all 625 projects for which information about benefit overrun and opening year are available.[9] For benefit overrun minus cost overrun, there is no significant movement over time, but the intercept stays highly significantly negative (BF=2352), indicating that benefit overrun is consistently less than cost overrun, similar to what we found above, and once more counter to Hirschman's claims. This is for the sample of 327 projects.[10][11]

In sum, the appraisal in this section shows there is no support in the available data for Hirschman's Benevolent Hiding Hand. Not only do the data not support Hirschman's main claim—that higher-than-estimated project costs will typically be outweighed by even higher-than-estimated benefits—the data show the exact opposite to be true: The typical (average) project is impeded by a double whammy of higher-than-estimated costs and lower-than-estimated benefits. This undermines project viability in a majority of cases instead of saving projects by the creative benefit-generation claimed by the Benevolent Hiding Hand. In other words, the Benevolent Hiding Hand is dominated by its evil twin, the Malevolent Hiding Hand. And not only do the data show this dominance to be remarkably consistent across all project types and geographies studied, it is also consistent over time, mainly due to deteriorating project performance on the benefit side, again in diametrical opposition to Hirschman's claims.

It should be stressed that the clear rejection of the Benevolent Hiding Hand above does not mean that projects do not exist for which it applies or that such projects may not be an interesting special case for study, as demonstrated by Sawyer (2014), who was a main inspiration for Hirschman.[12] Even in the dataset used above to reject the Benevolent Hiding Hand, it would be easy to fish out individual projects that confirm its basic idea. For instance, the German Karlsruhe-Bretten light-rail line, which is in the dataset, had a cost overrun in real terms of 78 percent but an even larger benefit overrun of 158 percent, making the project viable, in accordance with the Benevolent Hiding Hand. Similarly, the Danish Great Belt toll bridge—the longest suspension bridge in the world at the time of completion—had a cost overrun of 45 percent combined with a benefit overrun of 90 percent, again making the project fit Hirschman's claim. And so on. But to sample on the dependent variable like this and then call what you find a "general principle of action," as Hirschman (2015: 13) does, would be misleading.

The Oxford Advanced Learner's Dictionary defines a principle as "a general scientific theorem or law that has numerous special applications across a wide field." Following this definition, the Benevolent Hiding Hand cannot be said to constitute a principle; it applies to only 22 percent of cases and is thus too narrow in scope to count as a principle. It is a special case. Of the two Hiding Hands, only the Malevolent Hiding Hand applies widely enough—in 78 percent of cases, as documented above—to be called a principle. We conclude, therefore, that if the term principle is to be used in the context of the two Hiding Hands, like Hirschman did, it should be used for the Malevolent Hiding Hand, and for this only.

 

## Implications of Findings

The theoretical implications of our findings are clear. The idea of a Benevolent Hiding Hand is a special case and as an effort to capture reality, it is misleading or even a distraction. The Malevolent Hiding Hand is pervasive, and it is a case of the planning fallacy writ large—i.e., it applies not only to schedule, but also to costs and benefits in the widest sense—aggravated by the effects of ignorance, power, and motivated reasoning. The policy implications are equally clear. It is bad policy to justify plans and projects based on faith in the Benevolent Hiding Hand. In most cases initial costs and difficulties will not be overcome by later creativity and benefits; it is a dead-end at best, a scam at worst. Policy must reflect the reality of the Malevolent Hiding Hand, which hides obstacles and difficulties as well as systemic effects (Dorner, 1997), and must develop specific measures to overcome it to be effective (ibid.). Unbiased analysis of costs and benefits, and other technical tools (Sunstein and Hastie, 2015), count among those measures, and they have great promise, though there is a risk that such tools will themselves be infected by the Malevolent Hiding Hand. Ongoing and retrospective analysis can be helpful correctives.

We have emphasized the planning fallacy and noted that within behavioral science, it is well-understood that people typically underestimate the time that it takes to complete projects. This systematic error is partly a product of optimism bias (Sharot 2011); it also reflects motivated reasoning. When planners begin, they are likely to overlook obstacles or to believe that they can be surmounted. They might also pay too little attention to the secondary effects of interventions, which can raise new problems that must be separately addressed (Dorner 1997). The idea of the Hiding Hand can be seen, in part, as a generalization of the planning fallacy insofar as it rests on both unrealistic optimism and motivated reasoning. We have added that both of these may be accompanied or amplified by power; those who plan, or who favor plans, have a definite interest in downplaying obstacles. With his characteristic delight in paradoxes, Hirschman thought that the Hiding Hand was benevolent, because it would spur creativity and net benefits.

For development projects—Hirschman's topic—the implication is straightforward: The Benevolent Hiding Hand is itself a product of unrealistic optimism. Creativity does not regularly ride to the rescue, or provide unanticipatedly high benefits, offsetting unanticipatedly high costs. It follows that those who are responsible for new projects should beware of the planning fallacy, writ very large, and should be alert to the risk that costs will be higher, and benefits lower, than they anticipate.

We speculate that Hirschman's argument was deeply autobiographical, and part of its resonance stems from the fact that some version of the phenomenon can be seen in most people's lives. The Hiding Hand resonates with both romantics and cynics, which may well explain its lasting influence. Romantics are gratified by its intuitive appeal and its positive depiction of people as creative problem solvers who typically land their ventures on their feet, despite initial difficulties. Cynics see the theory as a means to justifying an end: getting projects started, and worrying about costs and benefits later. In the context of regulation, the Benevolent Hiding Hand does play a role, at least in the United States—in some important contexts, not because of unanticipatedly high gross benefits, but because of unanticipatedly low gross costs (Sunstein 2013). The basic cause here involves innovation: Confronted with regulatory requirements, companies are sometimes able to produce means of compliance that could not be anticipated at the time. This is not exactly Hirschman's mechanism, but it is extremely important. Nonetheless, the Malevolent Hiding Hand can also be found in the regulatory context (easily), and indeed it is about as common here as its more cheerful sibling (ibid.). Regulators are often too optimistic about costs, benefits, or both.

The Malevolent Hiding Hand is the planning fallacy writ large. This points to an important problem for welfare economics—and for any type of economics or policy that relies on cost-benefit analysis: the fact that ex ante estimates of costs and benefits can be erroneous or biased, as can be cost-benefit analysis, which might therefore prove a poor basis for decisionmaking. Our data show that an ex-ante benefit-cost ratio produced by conventional methods is typically overestimated by between 50 and 200 percent, depending on project type. In this context, ex-ante benefit-cost ratios are so misleading as to be worse than worthless, because decisionmakers might think they are being informed when in fact they are being misinformed. As a consequence, decisionmakers may give the green light to projects that should never have been started.

This does not prove the uselessness of cost-benefit analysis as such, needless to say. The task is to improve it, not to abandon it. In the American regulatory context, the track record seems to be far better, though existing information remains highly incomplete (Sunstein 2013). Existing studies, based on relatively small samples, find no systematic bias, though they do find a number of errors, both overestimating and underestimating net benefits (ibid.). We note that on this count, a great deal of further work remains to be done to justify firm conclusions, both in expanding the data set and in increasing peer review. The broader point is that if informed decisions are the goal, then conventional ex-ante cost-benefit analysis must be supplemented by unbiased technical advice, free from optimism bias, power play, and motivated reasoning, and by attempts to undertake projections in a way that fits with the best available evidence.

Economists have recently begun discussing the idea of "firing the forecaster," when forecasts are very wrong and the consequences severe (Akerlof and Shiller 2009, 146). We suggest, as a more general heuristic, to give forecasters skin in the game. Lawmakers and policymakers should develop institutional setups that reward forecasters who get their forecasts right and punish those who do not (if only through informal mechanisms for ex post approval and disapproval). We should not be surprised that forecasts are wrong if forecasters have no incentive to get them right. Only by creating such incentives will it be possible to begin to eliminate the worst consequences of the Malevolent Hiding Hand and the planning fallacy writ large.

Similarly, the technical analysis must be sharply separated from political motivations, so that the ex ante projection of costs and benefits is not distorted by those motivations. To promote accountability, the analysis should also be subject to external scrutiny (including scrutiny by independent experts and the public). With the caveats given above, the relatively good record of cost-benefit analysis in the American regulatory context, and the apparent absence of any systematic bias, is a product partly of technical competence, but also of a high degree of insulation from politics alongside accountability to the public.

## Conclusions

The idea of a Benevolent Hiding Hand offers an ingenious play on Smith's invisible hand, and it rightly suggests that human creativity is often underestimated, leading to unexpected solutions to seemingly intractable problems. In individual lives, as in public policy, blindness to obstacles sometimes has ironic and desirable consequences, precisely because it enables people to embark on projects that ultimately turn out well. What is true for some development projects is true for some regulations as well: Innovation drives cost down and benefits up.

It is an infectious account, but it does not fit the data. Far more often, planners are subject to the Malevolent Hiding Hand, which prompts people to proceed, unaware of the obstacles and of their inability to surmount them. The Hiding Hand obscures the planning fallacy, writ very large. Hirschman had a keen understanding of human psychology, but his enthusiasm for happy endings, and his delight in irony, led him to a misleading account of economic development. The Hiding Hand is usually malevolent.

                                                                                                      

*Table 1: Are cost overruns outweighed by even larger benefit overruns, as the Benevolent Hiding Hand would have it? The answer is a clear no.*

| Project type | Cost overrun | | Benefit overrun | | p* |
|---|---|---|---|---|---|
| | N | Average cost overrun (A/E) | N | Average benefit overrun (A/E) | |
| Dams | 243 | 1.96 | 84 | 0.89 | <0.0001 |
| BRT† | 6 | 1.41 | 4 | 0.42 | 0.007 |
| Rail | 264 | 1.40 | 74 | 0.66 | <0.0001 |
| Tunnels | 48 | 1.36 | 23 | 0.81 | 0.015 |
| Power plants | 100 | 1.36 | 23 | 0.94 | 00.0003 |
| Buildings | 24 | 1.36 | 20 | 0.99 | 0.01 |
| Bridges | 49 | 1.32 | 26 | 0.96 | <0.0001 |
| Roads | 869 | 1.24 | 532 | 0.96 | <0.0001 |
| **Total** | **1603** | **1.39 / 1.43‡** | **786** | **0.9 / 0.83‡** | **<0.0001** |

Sample of 2,062 projects; cost and benefit overruns measured as actual divided by estimated costs and benefits (A/E), respectively, in real terms.

*) The p-value of the test with null hypothesis that benefit overrun is actually larger than cost overrun, using Mann-Whitney test (smaller p-values are better).  †) Bus rapid transit.  ‡) Weighted and unweighted average, respectively.

*Table 2: Is the Benevolent Hiding Hand more prevalent than the Malevolent Hiding Hand? The answer is a clear no.*

| Project type | N | % of projects with Benevolent Hiding Hand (benefit overrun > cost overrun) | % of projects with Malevolent Hiding Hand (benefit overrun ≤ cost overrun) | p* |
|---|---|---|---|---|
| Dams | 78 | 18 | 82 | <0.0001 |
| BRT† | 4 | 0 | 100 | 0.06 |
| Rail | 48 | 17 | 83 | <0.0001 |
| Tunnels | 14 | 21 | 79 | 0.02 |
| Power plants | 23 | 26 | 74 | 0.017 |
| Buildings | 18 | 28 | 72 | 0.048 |
| Bridges | 16 | 38 | 62 | 0.22 |
| Roads | 126 | 25 | 75 | <0.0001 |
| **Total** | **327** | **22 / 22‡** | **78 / 78‡** | **<0.0001** |

*) The p-value of the test with null hypothesis of the Benevolent Hiding Hand, i.e., the number of projects with benefit overruns larger than cost overruns is actually greater than the number of projects with cost overruns larger than or equal to benefit overruns. Smaller p-values reject the Benevolent Hiding Hand. †) Bus rapid transit. ‡) Weighted and unweighted average, respectively. Note that the unweighted and weighted averages do not differ. Their p-values were different, but both were still below 0.0001, indicating high robustness of results.

**Notes**

[1] The closest we get to an assessment like this are Cracknell (1984) and Picciotto (1994a). Cracknell, who was an officer with the UK Overseas Development Administration (ODA), wrote that data from 200 evaluations of ODA projects "lend little support" to the Hiding Hand (Cracknell 1984: 17-18), but unfortunately Cracknell did not present data or analyses to substantiate his claim. Picciotto, who was an officer with the World Bank, tried to evaluate the Hiding Hand, but the evaluation lacks rigor and good data and weakly concludes, "the hiding hand has its advantages as well as disadvantages" (Picciotto 1994a: 223).

[2] Only two of Hirschman's projects are non-infrastructure, namely an industry project (the Karnaphuli paper mill mentioned in the main text) and a livestock project.

[3] Estimated costs and benefits are the estimates made at the time of decision to build (final business case). Actual costs are measured as recorded outturn costs; actual benefits as first-year benefits, or a later value as close to this as possible, if available and if first-year benefits were not available.

[4] It should be mentioned that results are probably conservative, i.e., cost overruns and benefit shortfalls in the project population are most likely larger than in the sample. This is because availability of data is often an indication of better-than-average project management, and because data from badly performing projects are often not released. This must be kept in mind when interpreting the results from statistical analyses, and it means that most likely the Hiding Hand is even more false in the project population than in the sample. For the full argument, see Flyvbjerg et al. (2002, 2005) and Flyvbjerg (2005).

[5] Significance is here defined in the conventional manner, with $p \leq 0.05$ being significant, $p \leq 0.01$ very significant, and $p \leq 0.001$ overwhelmingly significant.

[6] Ideally, data would be available for both cost overrun and benefit overrun for each project included in the statistical tests. However, data availability is far from ideal in the measurement of project performance. For only 327 projects out of the 2,062 in the sample were data available for both cost overrun and benefit overrun. Using this ideal criterion would therefore result in scrapping large amounts of useful information for the 1,735 other projects in the sample, which would clearly be unacceptable. We therefore decided to run the statistical tests twice, first for the 2,062 projects in the total sample, and second for the subsample of 327 projects with data available for both cost overrun and benefit overrun.

[7] Parameters for the models were estimated using MCMC. The language JAGS was used for this, through the rjags interface to R (Plummer 2003, 2012; R Core Team 2012). Statistical significance for these tests was measured by the Bayes Factor (BF) instead of by p-values, where $12 < BF \leq 150$ indicates a statistically significant result and $BF > 150$ indicates a highly significant result.

[8] When we analyze the smaller sample of 327 projects with available information for cost overrun, benefit overrun, and opening year, there appears to be a significant reduction in cost overrun over time (BF = 134) of about 0.5 percent per year. This sample includes data from 1952 to 2011. The larger sample of 1,271 projects is deemed to provide the more reliable results.

[9] When using the 327 projects with data for cost overrun, benefit overrun, and opening year, the significance disappears and we witness no movement over time, which again runs counter to the Benevolent Hiding Hand.

[10] For the period after 1980, the intercept is highly significantly negative at -1.031 (BF > 10000) but there is a significant movement over time: benefit overrun minus cost overrun is getting larger by 0.014 per year (BF=90). This is not a percentage increase, but a nominal yearly increase. However the increase is so small that, starting at the negative intercept, it would take 74 years for benefit overrun minus cost overrun to finally become positive, i.e., before the claim by Hirschman's Benevolent Hiding Hand would become true that higher-than-estimated costs are outweighed by even higher-than-estimated benefits. Finally, the positive trend is not supported by the larger and thus more informative samples of 1271 projects (cost overrun) and 625 projects (benefit overrun), respectively.

---

[11] Again, we tested for the influence of project type and geography using Bayesian modeling. Here we found statistical indication that cost overrun for dams have increased over time whereas cost overrun for rail has decreased; for the remaining six project types there was no statistically significant trend. For benefits, we found increasing overruns over time for power projects whereas overruns for roads were decreasing; again there was no statistically significant trend for the remaining six project types. Regarding geography, we found that cost overrun has increased over time in Latin America and North America, whereas cost overrun has decreased in Asia and Europe; for Africa and Oceania there were no statistically significant trends. We used the United Nation's macro-geographical (continental) regions as the basis for our geographical analyses, http://millenniumindicators.un.org/unsd/methods/m49/m49regin.htm. For benefit overrun, we found decreasing overruns (increasing benefit shortfalls) for Asia and Latin America; there was no statistically significant trend for Africa, Europe, North America, and Oceania. It should be mentioned that the differences between project types and geographies as regards change over time may be due to small numbers, especially for the early part of the period where observations are scant. Even if the dataset is the largest of its kind, when it is subdivided into eight project types, six regions, and up to nine time periods, some of the sub-samples become quite small and results correspondingly less reliable. But even with the small subsamples, when splitting by country and project type there is still a clear effect, which confirms just how strong that effect is.

[12] Marseille (1994) also describes a case where the Hiding Hand seems to apply.